\documentstyle[12pt,epsf]{article}
\newcommand{\dd}{\mbox{d}}
\newcommand{\eps}{\varepsilon}
\newcommand{\Li}{\mbox{Li}_2}
\newcommand{\vecc}[1]{\mbox{\boldmath $#1$}}

\def\la{\mathrel{\mathpalette\fun <}}

\def\fun#1#2{\lower3.6pt\vbox{\baselineskip0pt\lineskip.9pt
\ialign{$\mathsurround=0pt#1\hfil##\hfil$\crcr#2\crcr\sim\crcr}}}

\title{Large angle QED processes at $e^+e^-$ colliders
\\ at energies below 3 GeV
}

\author{A.B.~Arbuzov$^{1}$, G.V.~Fedotovich$^{2}$,
E.A.~Kuraev$^{1}$, N.P.~Merenkov$^{3}$, \\
V.D.~Rushai$^{4}$ ~and~ L.~Trentadue$^{5}$}

\date{}

\begin{document}

\maketitle

\begin{center}
{$^1$ \it Bogoliubov Laboratory of Theoretical Physics, JINR, \\
Dubna, 141980, Russia} \\[.2cm]
{$^2$ \it Budker Institute for Nuclear Physics, \\
Prospect Nauki, 11, Novosibirsk, 630090, Russia}\\[.2cm]
{$^3$ \it Kharkov Institute of Physics and Technology, \\
Kharkov, 310108, Ukraine} \\[.2cm]
{$^4$ \it Laboratory of Computing Techniques
and Automation, \\ JINR, Dubna, 141980, Russia}\\[.2cm]
{$^5$ \it Dipartimento di Fisica, Universit\'a di Parma and
INFN, \\ Gruppo Collegato di Parma, 43100 Parma, Italy}\\[1.2cm]
\end{center}

\begin{abstract}
QED processes at electron--positron colliders are
considerd. We present differential cross--sections for
large--angle Bhabha scattering, annihilation into muons
and photons. Radiative corrections
in the first order are taken into account exactly.
Leading logarithmic contributions are calculated in all
orders by means of the structure--function method.
An accuracy of the calculation can be estimated about 0.2\%.
\\[.2cm] \noindent
PACS~ 12.20.--m Quantum electrodynamics, 12.20.Ds Specific calculations
\end{abstract}

%
%

\section{Introduction}

At the existing VEPP--2M $e^+e^-$ collider
and planned meson factories QED processes
of the lowest order of the perturbation theory (PT) play
an important role. These processes are to be considered
as an essential background when extracting subtle mesons
properties from experimental data. QED processes are used also
for calibration and monitoring purposes. For instance, large--angle
Bhabha scattering is used for a precise
determination of luminosity at $e^+e^-$ colliders.
That is the reason, why radiative corrections (RC)
to QED processes are to be considered in detail.

A lot of attention was paid to the problem
of radiative correction calculations to various QED processes,
starting more than 50 years ago~\cite{old},
concerning mainly the lowest order of PT calculations.
The accuracy requirements of modern experiments, however, exceed
the ones provided by the first order RC. Unfortunately,
calculations of radiative corrections in higher orders encounter
tremendous technical difficulties. Nevertheless,
powerful methods developed in quantum chromodynamics
(see paper~\cite{QCD}
and references therein) provide a possibility
to improve essentially the results obtained earlier.
At first we mean the methods based on the renormalization group
ideas and on the factorization theorem. Thay allow us to
get a differential cross--section of a certain process similar
to the Drell--Yan process cross--section, and to consider leading
logarithmic RC of higher orders. Meanwhile, nonleading
contributions are to be taken from the lowest order PT calculations.

The aim of our paper
is to provide relevant theoretical formulae for QED processes, which
are required for experiments with CMD--2 and SND detectors
at VEPP--2M (Novosibirsk)~\cite{vepp2m} collider,
at DA$\Phi$NE (Frascati)~\cite{dafne}
and BEPC/BES (Beijing)~\cite{bepc} machines.
Formulae cited below may be applied also for higher energies
(see the Conclusions), if one will take into account weak interaction
and higher hadronic resonance contributions.

A cross--section calculated with an account of radiative
corrections (RC) in the $n$-th order of perturbation theory
(PT) contains enhanced contributions of the form $(\alpha/\pi)^nL^n$,
where $L=\ln(s/m_e^2)$ is the {\em large\/} logarithm
(for $s \sim 1$~GeV$^2$, $L\approx 15$). We call these
contributions the leading ones. Nonleading contributions
have the order $(\alpha/\pi)^nL^m$, $m<n$.
The terms, proportional to $(\alpha/\pi)^nL^n$, can be
calculated by means of the structure function method~\cite{strfun}.
The structure function formalism, based
on the renormalization group approach, permits one to keep all
leading terms of order $(\alpha L/\pi)^n$, $n=0,1,2\dots\ $
explicitly.

In the first order of PT the nonleading terms,
proportional to $(\alpha/\pi)$, can be accounted
by means of so--called ${\cal K}$--factor,
${\cal K}=1+(\alpha/\pi)K$.
As for nonleading terms of the order $(\alpha/\pi)^2L$, they
can be correctly calculated in the two--loop approximation. We will
consider them in further publications~\cite{twol}.
Fortunately, this second order nonleading RC are small:
$(\alpha/\pi)^2L \sim 10^{-4}$. So, the presented
formulae guarantee theoretical precision of the order $0.2\%$.
Considering hard photon emission in the first order of PT we distinguish
the contributions due to the initial state radiation, the final state
radiation, and their interference. The first one
always contains large logarithms. The ones due to the
final state radiation as well as the initial--final state
interference do not enhance the corrections considerably,
except the case of Bhabha scattering.
In the case of hard photon radiation large logarithms come
from kinematical regions, where the photona are emitted along
electron and positron beams. We will call
this kinematics as the collinear one.

Our work consist in explicit calculations of the Born and the first
order RC contributions to differential cross--sections.
We apply the structure function method to increase the accuracy.

The paper is organized as follows. In Section~2 we consider
$\mu^+\mu^-$ production. Both charge--even and charge--odd
contributions to the differential cross--section are evaluated.
In Sect.~3 we consider the electron--positron scattering process
at large angles. In Sect.~4 we investigate electron--positron
annihilation into photons. In the Conclusions we discuss the
results obtained and estimate the provided precision.
Numerical illustrations are given for a realistic experimental
set up.

\section{Muon Pair Production}

Consider the process
\begin{eqnarray*}
e^+(p_+) + e^-(p_-) \rightarrow \mu^+(q_+) + \mu^-(q_-).
\end{eqnarray*}
Taking into account the photon and $Z$--boson intermediate states,
the differential cross--section in the Born approximation
(in the framework of the Standard Model) has the form:
\begin{eqnarray}
\frac{\dd\sigma}{\dd\Omega}&=&\frac{\dd\sigma_0}{\dd\Omega}
\left\{1+K_{W} \right\}, \qquad
\frac{\dd\sigma_0}{\dd\Omega}=\frac{\alpha^2\beta}{4s}
\left(2-\beta^2(1-c^2)\right),
\\ \nonumber
\beta&=&\sqrt{1-4m_{\mu}^2/s},\qquad
c=\cos\widehat{\vecc{p}_-\vecc{q}}_-\, ,
\qquad s=(p_++p_-)^2 = 4\eps^2,
\end{eqnarray}
where the centre--of--mass reference system of the initial beams is
applied.

Here $K_W$ (we put the explicit expression for it in the Conclusions)
represents contributions due to $Z$--boson intermediate states
(see~\cite{born}
for example), we will neglect them within the accepted
precision: $K_W \sim s/M_Z^2 \la 10^{-3}$. We will drop also
all other contributions due to weak interactions in higher orders.

Consider first the even, with respect to the $c \leftrightarrow -c$
~permutation, part of the one--loop virtual and soft radiative
corrections.
Using the known in the literature Dirac and Pauli
muon form factors and the known soft photon contributions,
we obtain
\begin{eqnarray}
\frac{\dd\sigma^{B+S+V}_{\mathrm{even}}}{\dd\Omega} &=&
\frac{\dd\sigma_0}{\dd\Omega}\,\frac{1}{|1-\Pi(s)|^2}\Biggl\{1
+ \frac{2\alpha}{\pi}\Biggl[\biggl[L - 2
+ \frac{1+\beta^2}{2\beta}l_\beta \biggr]
\ln\frac{\Delta\eps}{\eps} \nonumber \\ \label{beven}
&+& \frac{3}{4}(L-1)
+ K_{\mathrm{even}} \Biggr]\Biggr\}, \\ \nonumber
K_{\mathrm{even}} &=& \frac{\pi^2}{6} - \frac{5}{4}
+ \rho\biggl(\frac{1+\beta^2}{2\beta} - \frac{1}{2}
+ \frac{1}{4\beta}\biggr) + \ln\frac{1+\beta}{2}\left(\frac{1}{2\beta}
+ \frac{1+\beta^2}{\beta}\right) \\ \nonumber
&-& \frac{1-\beta^2}{2\beta}\frac{l_\beta}{2-\beta^2(1-c^2)}
+ \frac{1+\beta^2}{2\beta}\biggl[ \frac{\pi^2}{6}
+ 2\Li\left(\frac{1-\beta}{1+\beta}\right)
+ \rho\ln\frac{1+\beta}{2\beta^2} \\ \nonumber
&+& 2\ln\frac{1+\beta}{2}\ln\frac{1+\beta}{2\beta^2} \biggr], \\ \nonumber
l_{\beta} &=& \ln\frac{1+\beta}{1-\beta}\, , \quad
\rho=\ln\frac{s}{m_{\mu}^2}\quad L=\ln\frac{s}{m_{e}^2}\, , \quad
\mbox{Li}_2(x)\equiv -\int\limits_{0}^{x}\frac{\dd t}{t}\ln(1-t).
\end{eqnarray}
where $\Delta\eps\ll\eps$ is the maximum energy of soft photon
in the centre--of--mass system.
$\Pi(s)$ is the vacuum polarization operator including
electron, muon, tau--meson and hadron contributions~\cite{vacpol}
(see Appendix~1).

The odd part of the one--loop correction comes from the interference
of Born and box Feynman diagrams and from the interference part
of the soft photon emission contribution.
It causes the charge asymmetry of the process:
\begin{equation}
\eta=[\dd\sigma(c)-\dd\sigma(-c)]/[\dd\sigma(c)+\dd\sigma(-c)]\neq 0.
\end{equation}
The odd part of the differential cross--section has the following form:
\begin{eqnarray} \label{eqll}
\frac{\dd\sigma^{S+V}_{\mathrm{odd}}}{\dd\Omega} &=&
\frac{\dd\sigma_0}{\dd\Omega}\,\frac{2\alpha}{\pi}\Biggl[
2\ln\frac{\Delta\eps}{\eps}\ln\frac{1-\beta c}{1+\beta c}
+ K_{\mathrm{odd}} \biggr], \\ \nonumber
K_{\mathrm{odd}} &=& \frac{1}{2}l_-^2 - L_-(\rho+l_-)
+ \mbox{Li}_2\left(\frac{1-\beta^2}{2(1-\beta c)}\right)
+ \mbox{Li}_2\left(\frac{\beta^2(1-c^2)}{1+\beta^2-2\beta c}\right)
\\ \nonumber
&-& \int\limits_{0}^{1-\beta^2}\frac{\dd x}{x}f(x)
\biggl(1-\frac{x(1+\beta^2-2\beta c)}{(1-\beta c)^2}\biggr)^{-\frac{1}{2}}
+ \frac{1}{2-\beta^2(1-c^2)}
\\ \nonumber
&\times& \Biggl\{ - \frac{1-2\beta^2+\beta^2c^2}
{1+\beta^2-2\beta c}(\rho+l_-)
- \frac{1}{4}(1-\beta^2)
\biggl[l_-^2 - 2L_-(l_-+\rho) \\ \nonumber
&+& 2\mbox{Li}_2\left(\frac{1-\beta^2}{2(1-\beta c)}\right)\biggr]
+ \beta c\biggl[-\frac{\rho}{2\beta^2} + \biggl(\frac{\pi^2}{12}
+ \frac{1}{4}\rho^2\biggr)\biggl(1- \frac{1}{\beta} - \frac{\beta}{2}
+ \frac{1}{2\beta^3}\biggr) \\ \nonumber
&+& \frac{1}{\beta}(-1-\frac{\beta^2}{2}
+ \frac{1}{2\beta^2}) \biggl(\rho\ln\frac{1+\beta}{2}
- 2\mbox{Li}_2\left(\frac{1-\beta}{2}\right)
- \mbox{Li}_2\left(-\frac{1-\beta}{1+\beta}\right)\biggr)
\\ \nonumber
&-& \frac{1}{2}l_-^2 + L_-(\rho+l_-)
- \mbox{Li}_2\left(\frac{1-\beta^2}{2(1-\beta c)}\right) \biggr]\Biggr\}
- (c\rightarrow -c), \\ \nonumber
f(x)&=&\biggl(\frac{1}{\sqrt{1-x}}-1\biggr)
\ln\frac{\sqrt{x}}{2} - \frac{1}{\sqrt{1-x}}\ln\frac{1+\sqrt{1-x}}{2}\, ,
\\ \nonumber
l_-&=&\ln\frac{1-\beta c}{2}\, ,\qquad
L_-=\ln\biggl(1-\frac{1-\beta^2}{2(1-\beta c)}\biggr).
\end{eqnarray}

In the ultra--relativistic limit $(\beta \to 1)$
for the even part we obtain:
\begin{eqnarray}
\left(\frac{\dd\sigma^{B+S+V}_{\mathrm{even}}}
{\dd\Omega}\right)_{\beta\to 1} &=&
\frac{\dd\sigma_0}{\dd\Omega}\,\frac{1}{|1-\Pi(s)|^2}\Biggl\{
1+\frac{2\alpha}{\pi}\biggl[(-2+L+\rho)\ln\frac{\Delta\eps}{\eps}
\nonumber \\
&-& 2 + \frac{3}{4}L + \frac{3}{4}\rho
+ \frac{\pi^2}{3} \biggr]\Biggr\}.
\end{eqnarray}
For the odd part in this limit we obtained the same
as Khriplovich~\cite{khrip}:
\begin{eqnarray}
\left(\frac{\dd\sigma^{S+V}_{\mathrm{odd}}}
{\dd\Omega}\right)_{\beta\to 1} &=&
\frac{\alpha^3}{s\pi}\Biggl\{ 2(1+c^2)\biggl[\ln(\mbox{ctg}\frac{\theta}{2})
\ln\frac{\eps}{\Delta\eps}
+ \frac{1}{2}\ln^2(\sin\frac{\theta}{2})
- \frac{1}{2}\ln^2(\cos\frac{\theta}{2}) \nonumber \\
&-& \frac{1}{4}\mbox{Li}_2(\sin^2\frac{\theta}{2})
+ \frac{1}{4}\mbox{Li}_2(\cos^2\frac{\theta}{2})\biggr]
+ \cos^2\frac{\theta}{2}\ln(\sin\frac{\theta}{2})
- \sin^2\frac{\theta}{2}\ln(\cos\frac{\theta}{2}) \nonumber \\
&-& \cos\theta\biggl[\ln^2(\cos\frac{\theta}{2})+\ln^2(\sin\frac{\theta}{2})
\biggr]\Biggr\}.
\end{eqnarray}

Consider now the process of hard photon emission
\begin{eqnarray}
e^+(p_+) + e^-(p_-) \rightarrow \mu^+(q_+) + \mu^-(q_-) + \gamma(k).
\end{eqnarray}
It was studied in detail~\cite{langle}.
The photon energy is assumed to be larger than $\Delta\eps$.
The differential cross--section has the form
\begin{eqnarray}
\dd\sigma&=&\frac{\alpha^3}{2\pi^2 s^2}R \dd\Gamma, \qquad
\dd\Gamma = \frac{\dd^3q_-\dd^3q_+\dd^3k}
{q_-^0q_+^0k^0}\delta^{(4)}(p_++p_--q_--q_+-k), \\ \nonumber
R&=&\frac{s}{16(4\pi\alpha)^3}\sum\limits_{spins}^{}
|M|^2=R_e + R_{\mu} + R_{e\mu}\, .  \label{rmu}
\end{eqnarray}
The contribution due to the initial state radiation reads
\begin{eqnarray}
R_e &=& A_e + B_e, \\ \nonumber
A_e &=& \frac{s}{s_1^2}\Biggl\{
\frac{s}{\chi_+\chi_-}
\left(\frac{1}{2}tt_1+\frac{1}{2}uu_1+sm^2_{\mu}\right)
+ \frac{1}{\chi_-}(2m_{\mu}^2\chi_+-u_1\chi_+'-t_1\chi_-') \\ \nonumber
&+& \frac{1}{\chi_+}(2m_{\mu}^2\chi_--u\chi_-'-t\chi'_+)
- \frac{1}{2}\biggl(\frac{u_1}{\chi_+}-\frac{t}{\chi_-}\biggr)(u-t_1)
- \frac{1}{2}\biggl(\frac{t_1}{\chi_+}-\frac{u}{\chi_-}\biggr)(t-u_1)
\\ \nonumber
&-& \frac{s}{2\chi_+}\left(2m_{\mu}^2
+ \frac{1}{\chi_-}(2m_{\mu}^2\chi_+-u_1\chi_+'-t_1\chi_-')\right) \\ \nonumber
&-& \frac{s}{2\chi_-}\left(2m_{\mu}^2
+ \frac{1}{\chi_+}(2m_{\mu}^2\chi_--u\chi_-'-t\chi_+')\right)
\Biggr\},
\\ \nonumber
B_e &=& - \frac{s}{s_1^2}\Biggl\{\left[\frac{m_e^2}{\chi_+^2}
+\frac{m_e^2}{\chi_-^2}\right]
\left(\frac{1}{2}tt_1+\frac{1}{2}uu_1+sm^2_{\mu}\right)
-\frac{m_e^2}{\chi_+^2}\bigl(2m_{\mu}^2\chi_-
- u\chi_-' - t\chi_+'\bigr)  \\ \nonumber
&-& \frac{m_e^2}{\chi_-^2}\bigl(2m_{\mu}^2\chi_+
- u_1\chi_+' - t_1\chi_-'\bigr)\biggr\},
\\ \nonumber
s&=&2p_+p_-,\quad s_1=(q_++q_-)^2, \quad t=-2p_-q_-, \quad t_1=-2p_+q_+,
\\ \nonumber
u&=&-2p_-q_+,\quad u_1=-2p_+q_-, \quad
\chi_{\pm}=p_{\pm}k, \quad \chi_{\pm}'=q_{\pm}k.
\end{eqnarray}
The final state radiation and the interference
of the initial and final state radiation contributions are
\begin{eqnarray}
R_{\mu}&=&A_{\mu}+B_{\mu}, \\ \nonumber
A_{\mu}&=&\frac{1}{s}\Biggl\{(tt_1+uu_1+2sm^2_{\mu})
\frac{q_-q_+}{\chi_+'\chi_-'}
- \frac{4m_{\mu}^2\chi_+\chi_-}{\chi_+'\chi_-'}
- \frac{t_1\chi_-+u\chi_+}{\chi_-'}
- \frac{u_1\chi_-+t\chi_+}{\chi_+'} \\ \nonumber
&+& \left(\frac{t_1}{2\chi_+'}-\frac{u_1}{2\chi_-'}\right)(t-u)
+\left(\frac{u}{2\chi_+'}-\frac{t}{2\chi_-'}\right)(u_1-t_1)  \\ \nonumber
&-& \frac{q_+q_-}{\chi_-'\chi_+'}\left[
(u_1+t_1)\chi_- + (u+t)\chi_+\right] \Biggr\}, \\ \nonumber
B_{\mu}&=&-\frac{tt_1+uu_1+2sm^2_{\mu}}{2s}\left(
\frac{m_{\mu}^2}{(\chi_+')^2}+\frac{m_{\mu}^2}{(\chi_-')^2}\right)
+ \frac{1}{s}\biggl(\frac{m_{\mu}^2}{(\chi_-')^2}(t_1\chi_-+u\chi_+)
\\ \nonumber
&+& \frac{m_{\mu}^2}{(\chi_+')^2}(u_1\chi_-+t\chi_+)\biggr),
\end{eqnarray}
\begin{eqnarray}
R_{e\mu}&=&-\frac{tt_1+uu_1+2sm^2_{\mu}}{2s_1}\left(
  \frac{t}{\chi_-\chi_-'}+\frac{t_1}{\chi_+\chi_+'}
- \frac{u}{\chi_-\chi_+'}-\frac{u_1}{\chi_+\chi_-'}\right)  \nonumber
\\ \nonumber
&-& \frac{2}{s_1}\biggl\{ - t - t_1
+ u + u_1 - \frac{1}{2}u_1(\frac{p_-}{\chi_-}-\frac{q_+}{\chi_+'})
(Q\chi_+'+P\chi_-) \\ \nonumber
&-& \frac{1}{2}t_1(\frac{p_-}{\chi_-}-\frac{q_-}{\chi_-'})
(Q\chi_-'-P\chi_-) - m_{\mu}^2(\chi_++\chi_-)(QP) \\ \nonumber
&+& \left(\frac{m_{\mu}^2}{\chi_+'}
- \frac{m_{\mu}^2}{\chi_-'}\right) (\chi_--\chi_+)
- \frac{1}{2}t(\frac{q_+}{\chi_+'}
- \frac{p_+}{\chi_+})(Q\chi_+'- P\chi_+) \\
&-& \frac{1}{2}u(\frac{q_-}{\chi_-'}-\frac{p_+}{\chi_+})
(Q\chi_-'+P\chi_+)\biggr\}, \\ \nonumber
P&=&\frac{p_+}{\chi_+}-\frac{p_-}{\chi_-}\, , \qquad
Q = \frac{q_-}{\chi_-'} - \frac{q_+}{\chi_+'}\, .
\end{eqnarray}
Quantity $R_e$ contains collinear and infrared
singularities. $A_{\mu}$ and $A_{e\mu}$ have only infrared
singularities. $B_{\mu}$ and $B_{e\mu}$ are free from
singularities. Quantity $R$ in the
ultra--relativistic case is given in Appendix~3.

After algebraic transformations we get
\begin{eqnarray}
R_e &=& \frac{s}{\chi_-\chi_+}B
- \frac{m_e^2}{2\chi_-^2}\;\frac{(t_1^2+u_1^2+2m_{\mu}^2s_1)}{s_1^2}
- \frac{m_e^2}{2\chi_+^2}\;\frac{(t^2+u^2+2m_{\mu}^2s_1)}{s_1^2}
+ \frac{m_{\mu}^2}{s_1^2} \Delta_{s_1s_1}\, ,
\nonumber \\
R_{e\mu} &=& B\biggl( \frac{u}{\chi_-\chi_+'} + \frac{u_1}{\chi_+\chi_-'}
- \frac{t}{\chi_-\chi_-'} - \frac{t_1}{\chi_+\chi_+'}\biggr)
+ \frac{m_{\mu}^2}{ss_1} \Delta_{ss_1}\, ,
\\ \nonumber
R_{\mu} &=& \frac{s_1}{\chi_-'\chi_+'}B + \frac{m_{\mu}^2}{s^2} \Delta_{ss}\, ,
\qquad B = \frac{u^2+u_1^2+t^2+t_1^2}{4ss_1}\, ,
\\ \nonumber
\Delta_{s_1s_1} &=& - \frac{(t+u)^2+(t_1+u_1)^2}{2\chi_-\chi_+}\, ,
\\ \nonumber
\Delta_{ss} &=& - \frac{u^2+t_1^2+2sm_{\mu}^2}{2(\chi_-')^2}
- \frac{u_1^2+t^2+2sm_{\mu}^2}{2(\chi_+')^2}
+ \frac{1}{\chi_-'\chi_+'}\bigl( ss_1 - s^2 + tu +t_1u_1 - 2sm_{\mu}^2\bigr),
\\ \nonumber
\Delta_{ss_1} &=& \frac{s+s_1}{2}\biggl( \frac{u}{\chi_-\chi_+'}
+ \frac{u_1}{\chi_+\chi_-'} - \frac{t}{\chi_-\chi_-'}
- \frac{t_1}{\chi_+\chi_+'} \biggr)
+ \frac{2(u-t_1)}{\chi_-'} + \frac{2(u_1-t)}{\chi_+'}\, .
\end{eqnarray}
Note that from these expressions one may in a moment obtain
the corresponding matrix element of the cross symmetrical
process $e^-\mu^+\to e^-\mu^+\gamma$.
To be rigorous, we have note that in the cross symmetrical
channel one has to take into account some additional terms
proportional to $m_e^2$. In our channel they can be shown as
follows:
\begin{eqnarray}
&& R_{\mu}\rightarrow R_{\mu} + \frac{m_e^2}{s^2}\Delta_{ss}', \nonumber \\
&& \Delta_{ss}' = \frac{2m_{\mu}^2 s_1}{\chi_-'\chi_+'}
+ \frac{t+u_1}{\chi_-'} + \frac{t_1+u}{\chi_+'}
+ \frac{4m_{\mu}^2}{\chi_-'} + \frac{4m_{\mu}^2}{\chi_+'}\, .
\end{eqnarray}
We checked the matrix element by a comparison
with the one used in {\tt FORTRAN} program~\cite{bardin},
describing electron--muon scattering.

The sum of the hard photon contribution, integrated
over the photon phase volume with the condition $k^0>\Delta\eps$,
and the contribution due to the soft and virtual photon emission
does not depend on the auxiliary parameter
$\Delta=\Delta\eps/\eps \ll 1$.
The main contribution, proportional to the large logarithm,
comes from the integration of $R_e$ in the case of collinear
kinematics of photon emission. For definiteness let us consider
the case when the photon moves close to the initial electron
direction:
\begin{eqnarray}
\widehat{\vecc{p}_-\vecc{k}}=\theta\leq\theta_0\ll 1,\quad
\theta_0\gg\frac{m_e}{\eps}.
\end{eqnarray}
Here we can use
\begin{eqnarray}
R_e\bigg|_{\vecc{k}\parallel\vecc{p}_-}
= \frac{s^2}{s_1^2}\left\{\frac{1+(1-x)^2}{x\chi_-}
- \frac{m_e^2}{\chi_-^2}(1-x)\right\}\frac{tt_1+uu_1+2sm_{\mu}^2}{2},
\end{eqnarray}
where $x$ is the energy fraction carried away by the emitted photon,
$x=k^0/\eps=1-s_1/s$.
Performing the integration over the photon emission angles,
we can present the corresponding part of the cross--section
(a similar contribution of the hard photon emission along
the positron is included below also) in the form
\begin{eqnarray} \label{eqcd}
&& \left(\frac{\dd\sigma}{\dd\Omega_-}\right)_{\mathrm{coll}}=C+D, \\ \nonumber
&& C=\frac{\alpha}{2\pi}\biggl(\ln\frac{s}{m_e^2}-1\biggr)
\int\limits_{\Delta}^{1}\!\!\dd x \frac{1+(1-x)^2}{x}
\biggl[\frac{\dd\tilde{\sigma}_0(1-x,1)}{\dd\Omega_-}
+ \frac{\dd\tilde{\sigma}_0(1,1-x)}{\dd\Omega_-}\biggr], \\ \nonumber
&& D=\frac{\alpha}{2\pi}\int\limits_{\Delta}^{1}\!\!\dd x \biggl\{x
+ \frac{1+(1-x)^2}{x}\ln\frac{\theta_0^2}{4}\biggr\}
\biggl[\frac{\dd\tilde{\sigma}_0(1-x,1)}{\dd\Omega_-}
+ \frac{\dd\tilde{\sigma}_0(1,1-x)}{\dd\Omega_-}\biggr],
\end{eqnarray}
where $\dd\tilde{\sigma}_0(1-x_1,1-x_2) / \dd\Omega_-$ ~is the so--called
{\em shifted\/} Born differential cross--section. It describes the process
$e^+(p_+(1-x_2)) + e^-(p_-(1-x_1)) \to \mu^+(q_+) + \mu^-(q_-)$,
\begin{eqnarray}
&& \frac{\dd\tilde{\sigma}_0(z_1,z_2)}{\dd\Omega_-}
= \frac{\alpha^2}{4s}\;\frac{y_1[z_1^2(Y_1-y_1c)^2
+ z_2^2(Y_1+y_1c)^2+8z_1z_2m_{\mu}^2/s]}
{z_1^3z_2^3[z_1+z_2-(z_1-z_2)cY_1/y_1]}\, , \\ \nonumber
&& y_{1,2}^2=Y_{1,2}^2-\frac{4m_{\mu}^2}{s},\quad
Y_{1,2}=\frac{q_{-,+}^0}{\eps}\, , \quad z_{1,2}=1-x_{1,2}\, .
\end{eqnarray}
Using the conservation laws
\begin{eqnarray} \label{conserv}
&& z_1+z_2=Y_1+Y_2,\qquad z_1-z_2=y_1c_- +y_2c_+, \\ \nonumber
&& y_1\sqrt{1-c_-^2}=y_2\sqrt{1-c_+^2},\quad c_-\equiv c, \quad
c_+=\cos\widehat{\vecc{p}_-\vecc{q}}_+\, ,
\end{eqnarray}
we obtain the energy fraction of the created muon
\begin{eqnarray}
Y_1&=& \frac{4m_{\mu}^2}{s}\;
\frac{(z_2-z_1)c}{2z_1z_2+[4z_1^2z_2^2-4(m_{\mu}^2/s)((z_1+z_2)^2
-(z_1-z_2)^2c^2)]^{1/2}} \nonumber \\
&+& \frac{2z_1z_2}{z_1+z_2-c(z_1-z_2)}\, .
\end{eqnarray}
Quantity $C$, after adding the corrections due to soft and virtual
photons, turns out to be the lowest order perturbative
expansion of the convolution
of the structure function ${\cal D}$
with the shifted Born differential cross--section.
Quantity $D$ plays role of a compensating term. Namely,
in the sum with the contribution of the cross--section
due to hard $(k^0 > \Delta\eps)$ photon emission at angles
(with respect to the electron and positron) larger than $\theta_0$
the dependence on the auxiliary parameters will cancel.

Here we remind about experimental conditions of the final particles
detection mentioned above. They are to be imposed explicitly by
introducing the restriction of the following kind:
\begin{eqnarray}
\Theta(z_1,z_2)=\Theta(Y_1-y_{\mathrm{th}})\Theta(Y_2-y_{\mathrm{th}})
\Theta(\cos^2\Psi_0-c_+^2)\Theta(\cos^2\Psi_0-c_-^2),\quad
c_+ = \widehat{\vecc{p}_+\vecc{q}}_+\, ,
\end{eqnarray}
where $y_{\mathrm{th}}\eps=\eps_{\mathrm{th}}$ is the threshold
of the detectors, angle
$\Psi_0$ determines the {\em dead\/} cones, surrounding beam
axes, unattainable for detection. More detailed cuts can be
implemented in a Monte Carlo program, using the formulae given
above.

The leading contributions to the cross--section, containing large
logarithm $L$, as may be recognized, combine to the kernel of
Altarelli--Parisi--Lipatov evolution equation:
\begin{eqnarray} \label{p12}
\dd\sigma &=& \int \dd z_1 \dd z_2 {\cal D}(z_1)
{\cal D}(z_2) \frac{\dd\tilde{\sigma}_0(z_1,z_2)}
{|1-\Pi(sz_1z_2)|^2}, \\ \nonumber
{\cal D}(z)&=&\delta(1-z)+\frac{\alpha}{2\pi}(L-1) P^{(1)}(z)
+\left(\frac{\alpha}{2\pi}\right)^2\frac{(L-1)^2}{2!}P^{(2)}(z)
+ \dots\ ,
\\ \nonumber
P^{(1)}(z)&=&\lim\limits_{\Delta \to 0}\left\{\delta(1-z)(2\ln\Delta
+ \frac{3}{2}) + \Theta(1-z-\Delta)\frac{1+z^2}{1-z}\right\}, \\ \nonumber
P^{(2)}(z) &=& \int\limits_{z}^{1}\frac{\dd t}{t}P^{(1)}(t)P^{(1)}
\left(\frac{z}{t}\right),\qquad
\int\limits_{0}^{1}\dd z P^{(1,2)}(z) = 0.
\end{eqnarray}
This formula is valid in the leading logarithmical approximation.
We will modify it by including nonleading contributions and using the
smoothed representation for structure functions~\cite{strfun}:
\begin{eqnarray}
{\cal D}(z,s)&=&{\cal D}^{\gamma}(z,s)+{\cal D}^{e^+e^-}(z,s), \\ \nonumber
{\cal D}^{\gamma}(z,s)&=&\frac{1}{2}b\biggl(1-z\biggr)^{\frac{b}{2}-1}
\biggl[1+\frac{3}{8}b + \frac{b^2}{16}\biggl(\frac{9}{8}
- \frac{\pi^2}{3}\biggr)\biggr] \\ \nonumber
&-& \frac{1}{4}b(1+z)
+ \frac{1}{32}b^2\biggl(4(1+z)\ln\frac{1}{1-z}
+ \frac{1+3z^2}{1-z}\ln\frac{1}{z} - 5 - z \biggr), \\ \nonumber
{\cal D}^{e^+e^-}(z,s)&=&\frac{1}{2}b\biggl(1-z\biggr)^{\frac{b}{2}-1}
\biggl[ - \frac{b^2}{288}(2L - 15) \biggr] \\ \nonumber
&+& \left(\frac{\alpha}{\pi}\right)^2\biggl[
\frac{1}{12(1-z)}\biggl(1-z-\frac{2m_e}{\eps}\biggr)^{\frac{b}{2}}
\biggl(\ln\frac{s(1-z)^2}{m_e^2}-\frac{5}{3}\biggr)^2 \\ \nonumber
&\times& \biggl(1+z^2+\frac{b}{6}\biggl(\ln\frac{s(1-z)^2}{m_e^2}
-\frac{5}{3}\biggr)\biggr) + \frac{1}{4}L^2\biggl(\frac{2}{3}\;
\frac{1-z^3}{z} + \frac{1}{2}(1-z) \\ \nonumber
&+& (1+z)\ln z\biggr) \biggr]
\Theta(1-z-\frac{2m_e}{\eps}), \qquad
b = \frac{2\alpha}{\pi}(L-1).
\end{eqnarray}
In comparison with the corresponding formula in ref.~\cite{strfun}
we shifted the terms, arising
due to virtual $e^+e^-$ pair production corrections, from ${\cal D}^{\gamma}$
into ${\cal D}^{e^+e^-}$.

Finally, the differential cross--section can be presented
in the form
\begin{eqnarray}
&& \frac{\dd\sigma^{e^+e^-\to\mu^+\mu^-(\gamma)}}{\dd \Omega_-}
=\int\limits_{z_{\mathrm{min}}}^{1}
\int\limits_{z_{\mathrm{min}}}^{1}\dd z_1\dd z_2\;
\frac{{\cal D}(z_1,s){\cal D}(z_2,s)
}{|1-\Pi(sz_1z_2)|^2}\, \frac{\dd\tilde{\sigma}_0(z_1,z_2)}{\dd \Omega_-}
\biggl(1+\frac{\alpha}{\pi}K\biggr)
\nonumber \\ \nonumber && \quad
+ \Biggl\{
\frac{\alpha^3}{2\pi^2s^2}\!\! \int\limits_{\stackrel{k^0>\Delta\eps}
{\widehat{k p}_{\pm}>\theta_0}}
\!\!\!\!\frac{R_e|_{m_e=0}}{|1-\Pi(s_1)|^2}\,\frac{\dd\Gamma}{\dd \Omega_-}
\; +\; \frac{D}{|1-\Pi(s_1)|^2}\Biggr\}
\\  && \quad
+ \Biggl\{\frac{\alpha^3}{2\pi^2s^2}\int\limits_{k^0>\Delta\eps}\!\!\!
\biggl(\mbox{Re}\, \frac{R_{e\mu}}{(1-\Pi(s_1))(1-\Pi(s))^{*}}
+ \frac{R_{\mu}}{|1-\Pi(s)|^2}\biggr)\frac{\dd\Gamma}{\dd \Omega_-} \nonumber \\  && \quad
+ \mbox{Re}\,\frac{C_{e\mu}}{(1-\Pi(s_1))(1-\Pi(s))^{*}}
+ \frac{C_{\mu}}{|1-\Pi(s)|^2} \Biggr\},                     \label{eemmg}
\\ && \nonumber
C_{\mu} = \frac{2\alpha}{\pi}\frac{\dd\sigma_0}{\dd\Omega_-}
\ln\frac{\Delta\eps}{\eps}\;
\biggl(\frac{1+\beta^2}{2\beta}\ln\frac{1+\beta}{1-\beta}-1\biggr),
\qquad z_{\mathrm{min}} = \frac{2m_{\mu}}{2\eps-m_{\mu}}\, ,
\\ \nonumber &&
C_{e\mu} = \frac{4\alpha}{\pi}\frac{\dd\sigma_0}{\dd\Omega_-}
\ln\frac{\Delta\eps}{\eps}\;
\ln\frac{1-\beta c}{1+\beta c}\, ,
\qquad
K=K_{\mathrm{odd}}+K_{\mathrm{even}},
\end{eqnarray}
where $D$, $C_{e\mu}$ and $C_{\mu}$ are compensating terms,
which provide cancellation of auxiliary parameters $\Delta$ and $\theta_0$
inside figure brackets. In the first term, containing ${\cal D}$ functions,
we gather all leading terms. A part of nonleading terms proportional
to the Born cross--section is written as the ${\cal K}$--factor. The rest
nonleading terms are written as two additional terms. The compensating
term $D$ (see Eq.~(\ref{eqcd})) comes from the integration in the collinear
region of hard photon emission. Quantities $C_{\mu}$ and $C_{e\mu}$
come from the even and odd parts of the differential cross--section
(arising due to soft and virtual corrections), respectively.
Here we consider the phase volumes of two
$(\dd\Omega_-)$ and three $(\dd\Gamma)$ final particles as
the ones, which already include all required experimental cuts.
Using the conservation laws Eq.~(\ref{conserv}) and concrete
experimental conditions one can define the lower
limits of the integration over $z_1$ and $z_2$.

There is a peculiar feature in the spectrum of hard photons.
Namely, in the end of the spectrum the differential cross--section
is proportional to the factor
\begin{eqnarray}
I(s_1) = \frac{2m_{\mu}^2+s_1}{s_1^2} \sqrt{1-\frac{4m_{\mu}^2}{s_1}}\, ,
\end{eqnarray}
which defines a peak at $s_1 \approx 5.6 m_{\mu}^2$.
It comes from the Feynman diagrams describing the emission by
the initial particles~\cite{bkf}.

\section{Large--Angle Bhabha Scattering}

The cross--section of Bhabha scattering (corrected
by the vacuum polarization factor), which enter into the
Drell--Yan form of corrected cross--section, has
a bit more complicated form, as far as the scattering
and annihilation amplitudes and their interference are
to be taken into account. We remind here the form of
the Lorentz--invariant matrix element module squared in the Born
approximation:
\begin{eqnarray}
R_0(s,t,u) &=& \frac{1}{16(4\pi\alpha)^4}\sum\limits_{\mathrm{spins}}
\left| {\cal M}(e^-(p_-) + e^+(p_+)\to e^-(p_-') + e^+(p_+'))\right|^2
\nonumber \\ \label{r0}
&=& \frac{s^2+u^2}{2t^2} + \frac{u^2+t^2}{2s^2} + \frac{u^2}{st}\, ,
\\ \nonumber
s &=& (p_-+p_+)^2, \quad t=(p_--p_-')^2, \quad u=(p_--p_+')^2,\quad
s+t+u = {\cal O}(m_e^2).
\end{eqnarray}
The first term in the right hand side describes the scattering--type
Feynman diagram square. The second one corresponds to the square of
the annihilation--type diagram. And the third one deals with the
interference of the two diagrams. A more compact representation of $R_0$
is also useful, $R_0 = (1\; +\; s/t\; +\; t/s)^2$.
The differential cross--section in the Born approximation has the form
\begin{eqnarray}
\frac{\dd{\sigma}^{\mathrm{Born}}_0}{\dd\Omega_-} = \frac{\alpha^2}{4s}
\left(\frac{3+c^2}{1-c}\right)^2.
\end{eqnarray}
We will need also quantity $R$ for arbitrary energies of initial particles.
Suppose that the initial electron and positron lost a certain energy
fraction. The corresponding kinematics is defined as follows:
\begin{eqnarray*}
&& e^-(z_1p_-)\; +\; e^+(z_2p_+)\; \longrightarrow \; e^-(\tilde{p}_-)\;
+\; e^+(\tilde{p}_+), \nonumber \\
&& \tilde{s}=sz_1z_2, \qquad \tilde{t}=-\frac{1}{2}sz_1Y_1(1-c), \qquad
\tilde{u} = - \frac{1}{2}sz_2Y_1(1+c), \nonumber \\
&& Y_1 =\frac{\tilde{p}_-^0}{\varepsilon} = \frac{2z_1z_2}{a},
\qquad a=z_1+z_2-(z_1-z_2)c.
\end{eqnarray*}
Here the {\em shifted\/} Born cross--section
corrected by vacuum polarization insertions into virtual photon
propagators reads
\begin{eqnarray}
\dd\tilde{\sigma}_{0}(z_1,z_2)
&=& \frac{4\alpha^2}{sa^2}\biggl\{\frac{1}{|1-\Pi(\tilde{t})|^2}\;
\frac{a^2+z_2^2(1+c)^2}{2z_1^2(1-c)^2}
+ \frac{1}{|1-\Pi(\tilde{s})|^2}\;
\frac{z_1^2(1-c)^2+z_2^2(1+c)^2}{2a^2} \nonumber \\ \label{shiftb}
&-& \mbox{Re}\;\frac{1}{(1-\Pi(\tilde{t}))(1-\Pi(\tilde{s}))^{*}}\;
\frac{z_2^2(1+c)^2}{az_1(1-c)} \biggr\}\dd\Omega_-\, .
\end{eqnarray}

Rewriting the known results~\cite{labs,labsr}
for the cross--section in the Born approximation
with one--loop virtual corrections to it and with the other
ones arising due to soft photon emission, we obtain
\begin{eqnarray}
\frac{\dd\sigma_{B+S+V}}{\dd\Omega_-}
&=& \frac{\dd\tilde{\sigma}_{0}(1,1)}{\dd\Omega_-}
\biggl\{1+ \frac{2\alpha}{\pi}(L-1)\left[2\ln\frac{\Delta\eps}{\eps}
+ \frac{3}{2}\right] \nonumber \\ \label{bsv}
&-& \frac{8\alpha}{\pi}\ln(\mbox{ctg}\frac{\theta}{2})
\ln\frac{\Delta\eps}{\eps} + \frac{\alpha}{\pi}K_{SV} \biggr\},
\end{eqnarray}
where
\begin{eqnarray}
K_{SV} &=& -1-2\Li(\sin^2\frac{\theta}{2}) + 2\Li(\cos^2\frac{\theta}{2})
+ \frac{1}{(3+c^2)^2}\biggl[\frac{\pi^2}{3}(2c^4 - 3c^3 - 15c)
\nonumber \\ \nonumber
&+& 2(2c^4 - 3c^3 + 9c^2 + 3c + 21)\ln^2(\sin\frac{\theta}{2})
- 4(c^4+c^2-2c)\ln^2(\cos\frac{\theta}{2}) \\ \nonumber
&-& 4(c^3+4c^2+5c+6)\ln^2(\mbox{tg}\frac{\theta}{2})
+ 2(c^3-3c^2+7c-5)\ln(\cos\frac{\theta}{2}) \\
&+& (\frac{10}{3}c^3+10c^2+2c+38)\ln(\sin\frac{\theta}{2}) \biggr]
\end{eqnarray}
is the part of the ${\cal K}$--factor coming from soft and virtual
photon corrections,
\begin{eqnarray}
\frac{\dd\tilde{\sigma}_{0}(1,1)}{\dd\Omega_-} &=& \frac{\alpha^2}{s}
\biggl\{\frac{5+2c+c^2}{2(1-c)^2|1-\Pi(t)|^2}
+ \frac{1+c^2}{4|1-\Pi(s)|^2}  \nonumber \\
&-& \mbox{Re}\;\frac{(1+c)^2}{2(1-c)(1-\Pi(t))(1-\Pi(s))^{*}}\biggr\},
\\ \nonumber
s &=& 4\eps^2,\quad t=-s\;\frac{1-c}{2}\, ,\quad
u=-s\;\frac{1+c}{2}\, , \quad
c = \cos\theta, \quad \theta=\widehat{\vecc{p}_-\vecc{p}}_-'\, .
\end{eqnarray}
Quantity $\Delta\eps$ in Eq.~(\ref{bsv}) is the maximum energy
of emitted soft photons. $\Pi(s)$ and $\Pi(t)$ are the vacuum
polarization operators in the $s$ and $t$ channels.
In the Conclusions we will estimate the contribution of weak interactions.


Consider now the process of hard photon (with the energy
$\omega =k^0 > \Delta\eps$) emission
\begin{eqnarray*}
e^+(p_+)\ +\ e^-(p_-)\ \to\ e^+(p_+')\ +\ e^-(p_-')\ +\ \gamma(k).
\end{eqnarray*}
We start with the differential cross--section in the form suggested
by F.A.~Berends et al. \cite{labs}
(which is valid for scattering angles being large compared
with $m_e/\eps$):
\begin{eqnarray} \label{bere}
\dd\sigma_{\mathrm{hard}} &=& \frac{\alpha^3}{2\pi^2s}\;
R_{e\bar{e}\gamma}\;\dd \Gamma , \qquad
\dd\Gamma = \frac{\dd^3p_+'\dd^3p_-'\dd^3k}{\eps_+'\eps_-'k^0}
\delta^{(4)}(p_++p_--p_+'-p_-'-k), \\ \nonumber
R_{e\bar{e}\gamma} &=& \frac{WT}{4}
- \frac{m_e^2}{(\chi_+')^2}\left(\frac{s}{t}+\frac{t}{s}+1\right)^2
- \frac{m_e^2}{(\chi_-')^2}\left(\frac{s}{t_1}+\frac{t_1}{s}+1\right)^2
\\ \nonumber
&-& \frac{m_e^2}{\chi_+^2}\left(\frac{s_1}{t}+\frac{t}{s_1}+1\right)^2
- \frac{m_e^2}{\chi_-^2}\left(\frac{s_1}{t_1}+\frac{t_1}{s_1}+1\right)^2,
\end{eqnarray}
where
\begin{eqnarray}
W &=& \frac{s}{\chi_+\chi_-} + \frac{s_1}{\chi_+'\chi_-'}
- \frac{t_1}{\chi_+'\chi_+} \nonumber
- \frac{t}{\chi_-'\chi_-}
+ \frac{u}{\chi_+'\chi_-} + \frac{u_1}{\chi_-'\chi_+}\, , \\ \nonumber
T &=& \frac{ss_1(s^2+s_1^2) + tt_1(t^2+t_1^2)+uu_1(u^2+u_1^2)}{ss_1tt_1}\, ,
\end{eqnarray}
and the invariants are defined as
\begin{eqnarray}
\nonumber
s &=& 2p_-p_+,\quad s_1=2p_-'p_+',\quad t=-2p_-p_-',\quad
t_1=-2p_+p_+', \\ \nonumber
u &=& -2p_-p_+',\quad u_1=-2p_+p_-',\quad \chi_{\pm}=kp_{\pm},\quad
\chi_{\pm}'=kp_{\pm}'.
\end{eqnarray}

It is convenient to extract the contribution of the collinear
kinematics. We do that for the following reasons. First, it is natural
to separate the region with very a sharp behaviour of the cross--section
and to consider it carefully. Second, we keep in mind the idea of
the leading logarithm factorization, which is valid in all orders of
the perturbation theory. We will evaluate the collinear kinematical
regions in two different ways.
The first one (the quasireal
electron approximation) is suitable for a generalization in order to
account higher order leading corrections by means of the structure
function method. In this way we will obtain below the leading logarithmic
contributions and the compensating terms, which will provide
the cancellation of auxiliary parameters.
The second one (the direct calculation) is more
rigorous, it can be used as a check of the first one.
We discuss it in detail in Appendix~2.

To obtain explicit formulae for compensators it is needed to
consider four kinematical regions corresponding to hard photon
emission inside narrow cones, surrounding the initial and final charged
particle momenta. The vertices of the cones are taken in the interaction
point. We introduce a small auxiliary parameter $\theta_0$,
it should obey the restriction
\begin{eqnarray} \label{thet0}
m_e/\sqrt{s} \ll \theta_0 \ll 1.
\end{eqnarray}
So, we define a collinear kinematical region, as the part
of the whole phase space, in which the hard photon is
emitted within the cone of $\theta_0$ polar angle with respect to
the direction of motion of one of the charged particles.

Using the method of quasireal electrons~\cite{quasi},
the matrix element ${\cal M}$
(squared and summed up over polarization states) of the process of hard
photon emission can be expressed through a {\em shifted\/}
matrix element of the process without
photon emission (see Eq.~(4) in~\cite{quasi}):
\begin{eqnarray}
\sum |{\cal M}(p_1,k,p_1',{\cal X})|^2 &=& 4\pi\alpha\left[
\frac{1+(1-x)^2}{x(1-x)}\;\frac{1}{kp_1} - \frac{m^2}{(kp_1)^2} \right]
\sum | {\cal M}_0(p_1-k,p_1',{\cal X})|^2,
\nonumber \\
\sum |{\cal M}(p_1,p_1',k,{\cal X})|^2 &=& 4\pi\alpha\left[
\frac{y^2+Y^2}{\omega Y}\;\frac{\eps}{kp_1'} - \frac{m^2}{(kp_1')^2} \right]
\sum | {\cal M}_0(p_1,p_1'+k,{\cal X})|^2, \\ \nonumber
x&=&\frac{\omega}{\eps}\, , \quad p_1^0=\eps , \quad
y=\frac{{p_1'}^0}{\eps}\, , \quad Y = x+y,
\end{eqnarray}
where ${\cal X}$ denotes the momenta of non--radiating incoming
and outgoing particles in a concrete process.
The integration over the phase volume of the emitted photon
inside the narrow cone, surrounding its parent charged particle momentum,
gives the following factors:
\begin{eqnarray}
&& \frac{4\alpha}{16\pi^2}\int\frac{\dd^3k}{\omega}\left[
\frac{1+(1-x)^2}{x(1-x)}\;\frac{1}{kp_1} - \frac{m^2}{(kp_1)^2}
\right] = \frac{\alpha}{2\pi}\;\frac{\dd z_1}{z_1}
\biggl[P_{\Theta}^{(1)}(z_1)\biggl(L-1+\ln\frac{\theta_0^2}{4}\biggr)
\nonumber \\ && \qquad
+ 1 - z_1 \biggr], \quad z_1 = 1-x,  \\
&& \frac{4\alpha}{16\pi^2}\int\frac{\dd^3k}{\omega}\left[
\frac{y^2+Y^2}{x Y}\;\frac{1}{kp_1'} - \frac{m^2}{(kp_1')^2} \right]
= \frac{\alpha}{2\pi}\;\frac{\dd z_3}{z_3}
\biggl[P_{\Theta}^{(1)}(z_3)\biggl(L-1+\ln\frac{\theta_0^2}{4}
 \nonumber \\ && \qquad \nonumber
+ 2\ln z_3\biggr)
+ 1 - z_3 \biggr], \qquad z_3 = 1 - \frac{\omega}{{p_1'}^0 + \omega}
= 1 - \frac{x}{Y}\, .
\end{eqnarray}
Note that the terms proportional to $(L - 1)$ contain the kernel $P^{(1)}$
(see Eq.~(\ref{p12}))
of Altarelli--Parisi--Lipatov evolution equations (more precisely, they
contain $\Theta$--part of the nonsinglet kernel):
\begin{eqnarray*}
P^{(1)}_{\Theta}(z) &=& \frac{1+z^2}{1-z}\Theta(1-z-\Delta).
\end{eqnarray*}
Collecting the contributions of the four collinear regions, we obtain
\begin{eqnarray}
\frac{\dd\sigma_{\mathrm{coll}}}{\dd\Omega_-}&=&\frac{\alpha}{\pi}
\int\limits_{\Delta}^{1}\frac{\dd x}{x}\; \Biggl\{
\biggl[\left(1-x+\frac{x^2}{2}\right)
\left(L - 1 + \ln\frac{\theta_0^2}{4} + 2\ln(1-x)\right)
+ \frac{x^2}{2}\biggr] \nonumber \\ &\times&  \nonumber
2\,\frac{\dd\tilde{\sigma}_{0}(1,1)}{\dd\Omega_-}
+ \biggl[\left(1-x+\frac{x^2}{2}\right)
\left(L - 1 + \ln\frac{\theta_0^2}{4}\right) + \frac{x^2}{2}\biggr]
\\ &\times&                                             \label{bhcoll}
\left[ \frac{\dd\tilde{\sigma}_{0}(1-x,1)}{\dd\Omega_-}
+ \frac{\dd\tilde{\sigma}_{0}(1,1-x)}{\dd\Omega_-} \right]
\Biggr\},
\end{eqnarray}
where the shifted Born cross--section is defined in Eq.~(\ref{shiftb}).

Adding the contributions of virtual and soft photon emission,
we restore the complete kernel. Generalizing the procedure
for the case of photon emission by all charged particles, we
come to the representation of the cross--section in the leading
logarithmic approximation. The final expression for the cross--section
therefore has the form
\begin{eqnarray}
&& \frac{\dd\sigma^{e^+e^-\to e^+e^-(\gamma)}}{\dd\Omega_-}
= \int\limits_{\bar{z}_1}^{1}\dd z_1\;\int\limits_{\bar{z}_2}^{1}\dd z_2\;
{\cal D}(z_1){\cal D}(z_2)
\frac{\dd\tilde{\sigma}_{0}(z_1,z_2)}{\dd\Omega_-}
\left(1+\frac{\alpha}{\pi}K_{SV}\right)\Theta \nonumber \\ \nonumber && \qquad
\times \int\limits^{Y_1}_{y_{\mathrm{th}}}\frac{\dd y_1}{Y_1}\;
\int\limits^{Y_2}_{y_{\mathrm{th}}}\frac{\dd y_2}{Y_2}\;
{\cal D}(\frac{y_1}{Y_1}){\cal D}(\frac{y_2}{Y_2})
\\ \nonumber
&& \qquad + \frac{\alpha}{\pi}\int\limits_{\Delta}^{1}\frac{\dd x}{x}
\Biggl\{\biggl[\left(1-x+\frac{x^2}{2}\right)\ln\frac{\theta_0^2(1-x)^2}{4}
+ \frac{x^2}{2}\biggr]\, 2 \,
\frac{\dd\sigma_0^{\mathrm{Born}}}{\dd \Omega_-}
\\ \nonumber && \qquad
+ \biggl[\left(1-x+\frac{x^2}{2}\right)\ln\frac{\theta_0^2}{4}
+ \frac{x^2}{2}\biggr]
\Biggl[\frac{4\alpha^2}{s(1-x)^2[2-x(1-c)]^4} \\ \nonumber && \qquad \times
\left(\frac{3-3x+x^2+2x(2-x)c+c^2(1-x+x^2)}{1-c}\right)^2
\\ \nonumber && \qquad
+ \frac{4\alpha^2}{s[2-x(1+c)]^4}
\left(\frac{3-3x+x^2-2x(2-x)c+c^2(1-x+x^2)}{1-c}\right)^2 \Biggr]
\Biggr\}\Theta
\\ && \qquad \label{bhabha}
- \frac{\alpha^2}{4s}
\left(\frac{3+c^2}{1-c}\right)^2
\frac{8\alpha}{\pi}\ln(\mbox{ctg}\frac{\theta}{2})
\ln\frac{\Delta\eps}{\eps}
\; + \; \frac{\alpha^3}{2\pi^2s}\!\!\!\!\!
\int\limits_{\stackrel{k^0>\Delta\eps}{\pi-\theta_0>\theta>\theta_0}}
\!\!\!\!\!\! \frac{WT}{4}\;\Theta \frac{\dd \Gamma}{\dd \Omega_-},
\\ \nonumber &&
Y_1 = \frac{2z_1z_2}{z_1+z_2-c(z_1-z_2)}\, , \qquad
Y_2 = \frac{z_1^2+z_2^2-(z_1^2-z_2^2)c}{z_1+z_2-c(z_1-z_2)}\, , \\ \nonumber &&
\bar{z}_1 = \frac{y_{\mathrm{th}}(1+c)}{2-y_{\mathrm{th}}(1-c)}\, ,\qquad
\bar{z}_2 = \frac{z_1y_{\mathrm{th}}(1-c)}{2z_1-y_{\mathrm{th}}(1+c)}\, .
\end{eqnarray}
The last term describes hard photon emission
process, provided that the photon energy fraction $x$ is larger than
$\Delta=\Delta\varepsilon/\varepsilon$, and its emission angle with respect
to any charged particle direction is larger than some small quantity
$\theta_0$. The sum of the last 3 terms in Eq.~(\ref{bhabha}) does not
depend on the auxiliary parameters $\Delta$ and $\theta_0$,
if they are sufficiently small.
We omitted the effects due to vacuum polarization in the last three
terms which describe real hard photon emission. Because the theoretical
uncertainty, coming from this approximation, has the order
$\delta(\dd\sigma)/\dd\sigma\sim (\alpha/\pi)^2L\la 10^{-4}$.
Nevertheless if the centre--off--mass energy is close to some
resonance mass (say to $m_{\phi}$) the effect due to vacuum
polarization may become visible. The differential
cross--section of non--collinear hard photon emission, that takes
into account vacuum polarization explicitly, is presented in Appendix~3.

\section{Annihilation of $e^+e^-$ into photons}

Considering the RC due to emission of virtual and soft real photons
to the cross--section of two quantum annihilation process
\begin{equation} \nonumber
e^+(p_+)\ +\ e^-(p_-)\ \to\ \gamma(q_1)\ +\ \gamma(q_2),
\end{equation}
we will use the results obtained in papers~\cite{eegg}:
\begin{eqnarray}
\dd\sigma_{B+S+V} &=& \dd\tilde{\sigma}_0(1,1)\Biggl\{1
+ \frac{\alpha}{\pi}
\biggl[(L-1)\biggl(2\ln\frac{\Delta\eps}{\eps}+\frac{3}{2}\biggr)
+K_{SV}\biggr]\Biggr\}, \\ \nonumber
K_{SV} &=& \frac{\pi^2}{3}+\frac{1-c^2}{2(1+c^2)}\biggl[\biggl(1
+ \frac{3}{2}\,\frac{1+c}{1-c}\biggr)\ln\frac{1-c}{2} \\ \nonumber
&+& \biggl(1+\frac{1-c}{1+c}+\frac{1}{2}\,\frac{1+c}{1-c}\biggr)
\ln^2\frac{1-c}{2}\;
+\; (c\to -c)\biggr], \\ \nonumber
\dd\tilde{\sigma}_0(1,1)
&=& \frac{\alpha^2(1+c^2)}{s(1-c^2)}\dd\Omega_1 \, , \quad
s=(p_++p_-)^2,\quad c=\cos\theta_1,\quad
\theta_1=\widehat{\vecc{q}_1\vecc{p}}_-\, .
\end{eqnarray}
We suppose that the two final photons are registered in an experiment
and their polar angles with respect to the initial beam directions are
not small ($\theta_{1,2}\gg m_e/\varepsilon$).

Consider the three--quantum annihilation process
\begin{eqnarray*}
e^+(p_+)\ +\ e^-(p_-)\ \to\ \gamma(q_1)\ +\ \gamma(q_2)\ +\ \gamma(q_3)
\end{eqnarray*}
with the cross--section (see the paper by M.V.~Terentjev~\cite{eegg})
\begin{eqnarray}
\dd\sigma^{e^+e^-\to 3\gamma} &=& \frac{\alpha^3}{8\pi^2 s}\;
R_{3\gamma}\,\dd\Gamma\, , \\ \nonumber
R_{3\gamma} &=& s\;\frac{\chi_3^2+(\chi_3')^2}{\chi_1\chi_2\chi_1'\chi_2'}
- 2m_e^2\biggl[\frac{\chi_1^2+\chi_2^2}{\chi_1\chi_2(\chi_3')^2}
+ \frac{(\chi_1')^2+(\chi_2')^2}{\chi_1'\chi_2'\chi_3^2}\biggr] \\ \nonumber
&+& \mbox{two cyclic permutations}, \\ \nonumber
\dd \Gamma &=& \frac{\dd^3q_1\dd^3q_2\dd^3q_3}{q_1^0q_2^0q_3^0}
\delta^{(4)}(p_++p_--q_1-q_2-q_3),
\end{eqnarray}
where
\begin{eqnarray*}
\chi_i=q_ip_-,\quad \chi_i'=q_ip_+,\quad i=1,2,3\, .
\end{eqnarray*}
The process can be treated as a radiative correction to
the two--quantum annihilation.

In the same way as we have done before we will distinguish the contributions
of the collinear kinematical region, when extra photons are emitted within
narrow cones of the opening angle $2\theta_0\ll 1$ to one of the charged
particles and the semi--collinear ones, when extra photons are emitted
outside these cones. This contribution can be obtained using the
quasireal electron method~\cite{quasi}.
It reads:
\begin{eqnarray}
\dd\sigma_{\mathrm{coll}}
&=& \frac{\alpha}{\pi}\int\limits_{\Delta}^{1}\frac{\dd x}{x}
\left[(1-x+\frac{x^2}{2})\biggl(L-1+\ln\frac{\theta^2_0}{4}\biggr)
+ \frac{x^2}{2}\right] \\ \nonumber &\times&
\left[\dd\tilde\sigma_0(1-x,1) + \dd\tilde\sigma_0(1,1-x)\right],
\end{eqnarray}
where the {\em shifted\/} cross--section has the form
\begin{eqnarray}
\dd\sigma_0(z_1,z_2)
= \frac{2\alpha^2}{s}\;\frac{z_1^2(1-c)^2+z_2^2(1+c)^2}
{(1-c^2)(z_1+z_2+(z_2-z_1)c)^2}\dd\Omega_1\, .
\end{eqnarray}
Again rearranging the separate contributions and applying the structure
functions method, we obtain the {\em improved\/} cross--section
\begin{eqnarray}
&& \dd\sigma^{e^+e^-\to\gamma\gamma(\gamma)}
= \int\limits_{\bar{z}_1}^{1}\dd z_1\; {\cal D}(z_1)
\int\limits_{\bar{z}_2}^{1}\dd z_2\; {\cal D}(z_2)
\dd\tilde{\sigma}_0(z_1,z_2)
\left(1+\frac{\alpha}{\pi}K_{SV}\right) \nonumber \\ \nonumber && \quad
+\ \frac{\alpha}{\pi}\int\limits_{\Delta}^{1}\frac{\dd x}{x}\;
\left[\left(1-x+\frac{x^2}{2}\right)\ln\frac{\theta^2_0}{4}
+\frac{x^2}{2}\right]
\biggl[\dd\tilde{\sigma}_0(1-x,1) +
\dd\tilde{\sigma}_0(1,1-x)\biggr] \\ && \quad
+\ \frac{1}{3}\!\!\!\!\int\limits_{\stackrel{z_i\geq\Delta}
{\pi-\theta_0\geq\theta_i\geq\theta_0}}\!\!
\frac{4\alpha^3}{\pi^2s^2}
\biggl[\frac{z_3^2(1+c_3^2)}{z_1^2z_2^2(1-c_1^2)(1-c_2^2)}
+ \mbox{two cyclic permutations}\biggr] \dd \Gamma, \\ \nonumber
&& z_i=\frac{q_i^0}{\eps},\quad c_i=\cos\theta_i,\quad
\theta_i=\widehat{\vecc{p}_-\vecc{q}}_i\, ,
\end{eqnarray}
where lower limits $\bar{z}_{1,2}$ are defined in Eq.~(\ref{bhabha}).
The multiplier $\frac{1}{3}$ in the last term takes into
account the identity of the final photons.
The sum of the last two terms does not
depend on $\Delta$ and $\theta_0$. Note that the annihilation process
is a pure QED one, hadronic contributions as well as weak interaction
effects are far beyond the required accuracy.

\section{Conclusions}

Thus we had considered the series of processes at electron--positron
colliders of moderately high energies. We presented differential
cross--sections to be integrated over concrete
experimental conditions. The formulae are good as for semi--analytical
integration, as well as for the creation of a Monte Carlo event
generator~\cite{MC}.
In a separate publication we are going to present analysis of
the effects of radiative corrections for the conditions
of VEPP--2M (Novosibirsk), DA$\Phi$NE (Frascati)
and BEPC/BES (Beijing).
The idea of our approach was to separate the contributions
due to $2\to 2$ like processes and $2\to 3$ like ones.
The compensating terms allow us to eliminate the dependence
on auxiliary parameters in both contributions separately.

Note that all presented formulae are valid only for large--angle
processes. Indeed, in the region of very small angles
$\theta \sim m_e/\varepsilon$
of final particles with respect to the beam directions
there are contributions of double logarithmic approximation [9].
These small angle regions give the main part of the total
cross--section. We suppose that this kinematics is rejected
by experimental cuts.

In the Table and Figures we present some results of numerical
calculations according to our formulae.
We suppose that a process--event implies detecting of
two final particles with the polar angles with respect to the beam axes
more than some value $\Psi_0$.
The energies of the particles have to exeed some experimental threshold
$\eps_{\mathrm{th}}$.
A cut--off on the acollinearity of the final particle momenta
is possible. But we switched off it in the computations.

In the Table we give the values of different RC contributions to
large--angle Bhabha scattering cross--section. Here we
switched off also the cut--off on the final particle energies
(we used only kinematical restrictions).
The contributions (see Eq.(\ref{bhabha})) are defined as follows:
\begin{eqnarray}
\sigma_{tot} \equiv \int\frac{\dd\sigma^{e^+e^-\to e^+e^-(\gamma)}}
{\dd\Omega_-}\,\dd\Omega_-
= \int\frac{\dd \sigma_0^{\mathrm{Born}}}{\dd\Omega_-}\,\dd\Omega_-
\biggl[ 1 + \frac{1}{100\%}\biggl(
\delta_{\mathrm{VP}} + \delta^{\mathrm{ini}}_{\mathrm{SF}}
+ \delta^{\mathrm{fin}}_{\mathrm{SF}} + \delta_{\mathrm{K}}
+ \delta_{\gamma} \biggr)\biggr],
\end{eqnarray}
where $\delta_{\mathrm{VP}}$ is due to the vacuum polarization
being included into the Born level diagrams;
$\delta^{\mathrm{ini(fin)}}_{\mathrm{SF}}$ is due to the
initial (final) state leading logarithmic corrections;
$\delta_{\mathrm{K}}$ shows the impact of the ${\cal K}$--factor;
$\delta_{\gamma}$ describes the contribution of one hard photon
emission at large angles. The effect due to the width of $\phi$ meson
is included as a part of vacuum polarization. It is small for the
given integrated cross--sections. But for the description of a
differential cross--section it is important, especially for
large scattering angles (see \cite{phi,eepipi}).

%
%
\begin{table}
\caption{The values of Bhabha cross--section and radiative
corrections to it.}
\begin{tabular}{|c|c|c|c|c|c|c|c|}
\hline
$\theta_{\pm}$ & $\sigma^{\mathrm{Born}}_0$ (mb)
& $\delta_{\mathrm{VP}}$ (\%)
&$\delta^{\mathrm{ini}}_{\mathrm{SF}}$ (\%)
&$\delta^{\mathrm{fin}}_{\mathrm{SF}}$ (\%)
&$\delta_{\mathrm{K}}$ (\%) & $\delta_{\gamma}$ (\%)
& $\sigma_{\mathrm{tot}}$ (mb)   \\ \hline
$9^{\circ} < \theta_{\pm} < 171^{\circ}$
& $3.77\cdot 10^{-2}$ & 1.53 & 0.11 & - 0.30 & -1.82 & 1.50
& $3.81\cdot 10^{-2}$
\\ \hline
$1^{\circ} < \theta_{\pm} < 179^{\circ}$
& $3.08$ & 0.80 & 0.70 & -0.17 & - 3.5 & 4.74 & 3.16
\\ \hline 
\end{tabular}
\end{table}

In Figures~1 and~2 we illustrate the charge--odd part of the differential
cross--section for the $e^+e^-\to\mu^+\mu^-$ process (see Eq.(\ref{eemmg})).
The quantity
\begin{eqnarray}
A_{FB}=\frac{\dd\sigma^{e^+e^-\to\mu^+\mu^-(\gamma)}_{\mathrm{odd}}/\dd c}
{\dd\sigma^{e^+e^-\to\mu^+\mu^-(\gamma)}_{0}/\dd c}\, 100\%
\end{eqnarray}
is shown there as a function of $c$.
The short--dashed line represents the contribution due to soft photon
emission and virtual corrections. The long--dashed line represents the
corresponding odd contribution due to hard photon emission. It
comes from the intereference of the amplitudes due to initial
and final radiation $(R_{e\mu})$. In the sum of the two contributions
the dependence on the auxiliary parameter $\Delta$ disappears
(it was chosen $\Delta=0.01$). And we
obtain an experimentally measurable asymmetry $A_{FB}$ (the solid line).
There is also a contribution to the asymmetry due to electroweak
interactions.
Namely, due to the interference of the Born level amplitudes
with $\gamma$ and $Z$ boson in the $s$--channel. It can be found from
the weak ${\cal K}$--factor~(\ref{kweemm}). It is included in the
total sums (solid lines). But for the chosen energies it is really small
(it gives a maximal shift of about 0.01\% for $E_{\mathrm{beam}}=0.51$~GeV
and about 0.1\% for $E_{\mathrm{beam}}=1.55$~GeV.

Let us discuss the accuracy, provided by our formulae.
The contribution of weak interaction to the cross--section of muon
pair production and Bhabha scattering was parameterized be so--called
{\em weak\/} ${\cal K}$--factors:
\begin{eqnarray}
{\cal K}_{W} = \frac{(\dd\sigma)_{\mathrm{EW}}-(\dd\sigma)_{\mathrm{QED}}}
{(\dd\sigma)_{\mathrm{QED}}}\, ,
\end{eqnarray}
where quantities $(\dd\sigma)_{\mathrm{EW}}$ and $(\dd\sigma)_{\mathrm{QED}}$
are the cross--sections calculated in the Born approximation in the frames
of the Standard Model and QED, respectively. The weak ${\cal K}$--factors
(we used the results of papers~\cite{born,bornb}) are
\begin{eqnarray}
K_W^{e\bar{e}\to\mu\bar{\mu}} &=&
\frac{s^2(2-\beta^2(1-c^2))^{-1}}
{(s-M_Z^2)^2+M_Z^2\Gamma_Z^2}\biggl\{(2-\beta^2(1-c^2))
\left(c_v^2\biggl(3-2\frac{M_Z^2}{s}\biggr)+c_a^2\right) \nonumber \\
                                           \label{kweemm}
&-& \frac{1-\beta^2}{2}(c_a^2+c_v^2)
+c\beta\left[4\biggl(1-\frac{M_Z^2}{s}\biggr)c_a^2+8c_a^2c_v^2\right]\biggr\}, \\
c_a&=&-\frac{1}{2\sin 2\theta_W},\quad
c_v=c_a(1-4\sin^2\theta_W),
\nonumber \\
K_W^{e\bar{e}\to e\bar{e}}    &=&
\frac{(1-c)^2}{2(3+c^2)^2}\biggl[ 4B_1 + (1-c)^2B_2 + (1+c)^2B_3 \biggr] - 1,
\end{eqnarray}
\begin{eqnarray*}
B_1 &=& (\frac{s}{t})^2 \left|1+(g^2_v-g^2_a)\xi \right|^2  ,
\qquad B_2= \left| 1+(g^2_v-g^2_a) \chi \right|^2 ,       \\
B_3 &=&\frac{1}{2} \left| 1+\frac{s}{t}+(g_v+g_a)^2(\frac{s}{t}
\xi +\chi)\right|^2
+ \frac{1}{2} \left| 1+\frac{s}{t}+(g_v-g_a)^2(\frac{s}{t}
\xi +\chi)\right|^2, \\
\chi &=& \frac{\Lambda s}{s-m^2_z+iM_Z \Gamma _Z}\, , \qquad
\xi = \frac{\Lambda t}{t-M^2_Z}\, ,
\\ \Lambda &=& \frac{G_FM^2_Z}{2\sqrt 2 \pi \alpha}=(\sin 2\theta_W)^{-2},
\quad g_a=-\frac{1}{2},\quad
g_v=-\frac{1}{2} (1-4 \sin^2\theta_W),
\end{eqnarray*}
here $\theta_W$ is the weak mixing angle.

These quantities are of order 0.1\% up to $\sqrt{s} < 3$~GeV.
Contribution of weak interactions to the cross--section
of the annihilation into photons (which is absent at the Born level)
can be estimated as
\begin{eqnarray}
({\cal K}_{W})_{e\bar{e}\to \gamma\gamma} \la \frac{\alpha s}{\pi M_W^2}\, .
\end{eqnarray}
It comes from one--loop electroweak radiative corrections.
Another source of uncertainties comes from the approximation of
collinear kinematics (or the approximation of quasireal
electrons~\cite{quasi})
It can be estimated by the largest omitted terms
\begin{eqnarray}
\frac{\alpha}{\pi}\theta_0^2 \quad \mbox{and} \quad \frac{\alpha}{\pi}
\left(\frac{m_e}{\eps\theta_0}\right)^2.
\end{eqnarray}
\noindent
Really in the calculations we used the value of the parameter $\theta_0$
of the order $10^{-2}$ because of the restrictions $\theta_0\ll 1$ and
$\eps\theta_0/m_e \gg 1$. Note that the coefficient before terms of that sort
(calculable in principle) is the function of energy fractions and angles,
they are of order of 1. For typical values of energy $\eps =0.5$~GeV
this uncertainty is of order $2\cdot 10^{-4}$ or less.

The third source of uncertainties is the error in the definition of
the hadronic vacuum polarization. It has been estimated~\cite{vacpol}
to be of order 0.04\%. For $\phi$--meson factories a systematic error
in the definition of the $\phi$--meson contribution into vacuum
polarization is to be added.

Next point concerns nonleading terms of order $(\alpha/\pi)^2L$.
There are several sources of them. One is the emission of two extra
hard particles (for the case of Bhabha scattering it was considered
in the series of papers~\cite{labs}).
Other are related to virtual and soft--photon
radiative corrections to single hard photon emission and Born processes.
The most part of these contributions was not considered up to now.
Nevertheless, we can estimate the coefficient before the quantity
$(\alpha/\pi)^2L\approx 1\cdot 10^{-4}$ to be of order of unity.
That was indirectly confirmed by our complete calculations of
these terms for the case of small--angle Bhabha scattering~\cite{long}.

Considering all mentioned above sources of uncertainties as
independent, we conclude that the systematic error of our formulae
does not exceed 0.2\%. The main error is due to unknown
second--order next--to--leading radiative corrections.

For precise luminosity measurements we suggest to use the large--angle
Bhabha scattering process. It has a very large cross--section, a good
signature in detectors, and the lowest theoretical uncertainty.

\subsection*{Acknowledgement}
The authors are grateful to A.~Sher, S.~Panov and V.~Astakhov
for their close interest in the initial stage~\cite{prep},
to V.~Fadin and L.~Lipatov for critical comments,
and to D.~Bardin for cross checks.
This work was partially supported by INTAS grant 1867--93
and by RFBR grant $N^{\underline{\circ}}$~96--02--17512.
One of us (A.B.A.) is thankful to the INTAS foundation
for the financial support via the International Centre for
Fundamental Physics in Moscow.

\section*{Appendix~1}

We present here leptonic and hadronic contributions
into the vacuum polarization operator:
\begin{eqnarray}
\Pi &=& \Pi_l + \Pi_{h}, \\ \nonumber
\Pi_l(s) &=& \frac{\alpha}{\pi}\Pi_1(s)
+ \left(\frac{\alpha}{\pi}\right)^2\Pi_2(s)
+ \left(\frac{\alpha}{\pi}\right)^3\Pi_3(s) + \dots \\ \nonumber
\Pi_h(s) &=& \frac{s}{4\pi\alpha}\Biggl[
\mbox{PV}\int\limits_{4m_{\pi}^2}^{\infty}
\frac{\sigma^{e^+e^-\to\mathrm{hadrons}}(s')}{s'-s}\dd s'
- \mbox{i}\pi\sigma^{e^+e^-\to\mathrm{hadrons}}(s)\Biggr].
\end{eqnarray}
The first order leptonic contribution is well known~\cite{old}:
\begin{eqnarray}
\Pi_1(s) &=& \frac{1}{3}L - \frac{5}{9} + f(x_{\mu}) + f(x_{\tau})
- \mbox{i}\pi\left[\frac{1}{3} + \phi(x_{\mu})\Theta(1-x_{\mu})
+ \phi(x_{\tau})\Theta(1-x_{\tau})\right],
\end{eqnarray}
where
\begin{eqnarray*}
f(x) &=& \left\{\begin{array}{l}
-\frac{5}{9}-\frac{x}{3}+\frac{1}{6}(2+x)\sqrt{1-x}\ln\left|
\frac{1+\sqrt{1-x}}{1-\sqrt{1-x}}\right|\ \ \ \mbox{for}\ \ x\leq 1, \\
-\frac{5}{9}-\frac{x}{3}+\frac{1}{6}(2+x)\sqrt{1-x}\;\mbox{arctg}\left(
\frac{1}{\sqrt{x-1}}\right)\ \ \ \mbox{for}\ \  x > 1,\\
\end{array}\right. \\
\phi(x) &=& \frac{1}{6}(2+x)\sqrt{1-x},\qquad
x_{\mu,\tau} = \frac{4m_{\mu,\tau}^2}{s}\, .
\end{eqnarray*}
In the second order it is enough to take only the logarithmic term
from the electron contribution
\begin{eqnarray}
\Pi_2(s) = \frac{1}{4}(L-\mbox{i}\pi) + \zeta(3) - \frac{5}{24}\, .
\end{eqnarray}

A discussion of the $\phi$--meson contribution to the vacuum
polarization operator, which is important for $\sqrt{s}\approx m_{\phi}$,
can be found in~\cite{phi,eepipi}.

\section*{Appendix~2}

Here we present the direct evaluation of the collinear region
contribution to the Bhabha scattering cross--section.
Let us write the contribution of the collinear kinematics
in the form:
\begin{eqnarray}
(\dd \sigma )_{\mathrm{coll}} =
\dd \sigma_{\vecc{k}\parallel\vecc{p}_-}
+ \dd \sigma_{\vecc{k}\parallel\vecc{p}_+}
+ \dd \sigma_{\vecc{k}\parallel\vecc{p}_-'}
+ \dd \sigma_{\vecc{k}\parallel\vecc{p}_+'}
\equiv \dd\sigma_a + \dd\sigma_b + \dd\sigma_c + \dd\sigma_d\, .
\end{eqnarray}
For the case of photon emission along the initial electron we have
\begin{eqnarray}
W_a &=& \frac{2}{\omega^2}\;\frac{1}{1-\beta c_2},\qquad
T_a = \frac{1+(1-x)^2}{1-x}R(s_1,t^a,u^a),\qquad
\dd \Gamma_a = \frac{\dd^3k}{\omega}\,\frac{y_1^a}{a_a}
\dd\Omega_-, \\ \nonumber
c_2 &=& \cos(\widehat{\vecc{k}\vecc{p}}_-),\quad
\beta = \sqrt{1-\frac{m_e^2}{\varepsilon^2}}\, ,\quad \omega=k^0=x\eps, \quad
s_1 = s(1-x), \quad t_1^a=t^a(1-x),\\ \nonumber
u_1^a &=& u^a(1-x), \quad
t^a = -s\frac{(1-x)^2(1-c)}{a_a}\, , \quad
u^a = -s\frac{(1-x)(1+c)}{a_a}\, , \quad s=4\eps^2, \\ \nonumber
a_a &=& 2 - x(1-c),\quad y_1^a = \frac{2(1-x)}{a_a}\, ,
\quad c = \cos(\widehat{\vecc{p}_-\vecc{p}}_-').
\end{eqnarray}
Performing the angular integration over photon angles inside
the narrow cone, surrounding the direction of the initial electron beam,
we get
\begin{eqnarray}
\int W_a \dd\Gamma_a = 4\pi\frac{\dd\omega}{\omega}\dd\Omega_-
\int\limits_{1-\theta_0^2/2}^{1}\frac{\dd c_1}{1-\beta c_1} =
4\pi\frac{\dd x}{x}\dd\Omega_-\frac{y_1^a}{a_a^2}
\left(L + \ln\frac{\theta_0^2}{4}\right) + {\cal O}(\theta_0^2).
\end{eqnarray}
We neglect the terms proportional to $\theta_0^2$.
Collecting all the factors and reminding the contribution of
the terms proportional to $m_e^2$ (see Eq.~(\ref{bere})), we obtain
the contribution of the first collinear region:
\begin{eqnarray}
\frac{\dd \sigma_a}{\dd\Omega_-} &=& \frac{4\alpha^2}{s}\;\frac{\alpha}{\pi}
\int\limits_{\Delta}^{1}\frac{\dd x}{x}\left[\left(1-x+\frac{x^2}{2}\right)
\left(L - 1 + \ln\frac{\theta^2}{4}\right) + \frac{x^2}{2}\right]
\frac{1}{a_a^2} \nonumber \\ &\times&
\frac{[a_a^2+(1-c)^2(1-x)^2-a_a(1-c)(1-x)]^2}{a_a^2
(1-x)^2(1-c)^2}\, .      \label{dsa}
\end{eqnarray}
For the case of photon emission inside the narrow cone,
surrounding $\vecc{p}_+$, in a similar way one gets
\begin{eqnarray}
\frac{\dd \sigma_b}{\dd\Omega_-} &=& \frac{4\alpha^2}{s}\;\frac{\alpha}{\pi}
\int\limits_{\Delta}^{1}\frac{\dd x}{x}\left[\left(1-x+\frac{x^2}{2}\right)
\left(L - 1 + \ln\frac{\theta^2}{4}\right) + \frac{x^2}{2}\right]
\frac{1}{a_b^2} \nonumber \\ &\times&
\frac{[a_b^2+(1-c)^2(1-x)^2-a_b(1-c)(1-x)]^2}{a_b^2(1-c)^2}\, , \qquad
a_b = 2 - x(1+c).         \label{dsb}
\end{eqnarray}
Here we used the following formulae:
\begin{eqnarray*}
W_b &=& \frac{2}{\omega^2}\;\frac{1}{1-\beta c_+}\, , \quad
T_b = \frac{1+(1-x)^2}{1-x}R(s_1,t^b,u^b),\quad
\dd \Gamma_b = \frac{\dd^3k}{\omega}\,\frac{y_1^b}{a_b}
\dd\Omega_-, \\ \nonumber
y_1^b &=& \frac{p_-^0}{\eps} = \frac{1-x}{a_b}\, , \quad
t^b = - s_1\frac{1-c}{a_b}\, , \quad u_b = - s_1\frac{(1+c)(1-x)}{a_b}\, ,\quad
t^b + u^b + s_1 = 0.
\end{eqnarray*}
For the cases $\vecc{k}\parallel\vecc{p}_-'$ and $\vecc{k}\parallel\vecc{p}_+'$
quantity R (if suppose $\Pi$=0) is simple:
\begin{eqnarray*}
R_c &=& R_d = \frac{1}{4}\left(\frac{3+c^2}{1-c}\right)^2, \qquad
T_c = T_d = \frac{1+(1-x)^2}{1-x}R_c, \\
\dd\Gamma_{c,d} &=& \varepsilon^2\dd x\;\dd\Omega_-\;\dd\phi_1\;\dd c_1\;
\frac{xy^{c,d}}{2-x+xc_1}\, , \quad c_1=\cos\widehat{\vecc{k}\vecc{p}_-'}, \\
y^{c} &=& \left. \frac{p_-'^0}{\varepsilon}
\right|_{\vecc{k}\parallel\vecc{p}_-'} \approx \left.\frac{2(1-x)}{2-x+xc_1}
\right|_{c_1\to 1} = 1-x, \\
y^{d} &=& \left. \frac{p_-'^0}{\varepsilon}
\right|_{\vecc{k}\parallel\vecc{p}_+'} \approx
\frac{2(1-x)}{2-x+xc_1} + c_1\frac{m^2_e}{4\varepsilon^2}\,\frac{x}{1-x}\, ,\\
W_{c}\cdot(kp_-') &=& W_{d}\cdot(kp_+') = \frac{2(1-x)}{x}\, .
\end{eqnarray*}
Note that $kp_+'=2\varepsilon^2(1-y)$.
In the evaluation of the rest multipliers for these cases one has to
be careful:
\begin{eqnarray*}
\left.\int W_c\dd\Gamma_c\right|_{1-\theta_0^2/2\leq c_1\leq 1}
&=& \left.\int W_d\dd\Gamma_d
\right|_{-1+\theta_0^2(1-x)^2/2\geq c_1\geq -1} \\
&=& 2\pi\frac{\dd x}{x}\dd\Omega_-\frac{1-x}{2}
\left[L + \ln\frac{\theta_0^2}{4} + 2\ln(1-x) \right].
\end{eqnarray*}
Note that the collinear region $d$ is defined by the condition
$1-\theta_0^2/2\leq \cos\widehat{\vecc{k}\vecc{p}_+'}\leq 1$, which
leads to the bounds on $c_1$ shown above.
So, the contributions of these two collinear regions are
\begin{eqnarray}
\frac{\dd\sigma_c+\dd\sigma_d}{\dd\Omega_-} &=& 2\,\frac{\alpha^2}{4s}
\left(\frac{3+c^2}{1-c}\right)^2\;\frac{\alpha}{\pi}\int\limits^{1}_{\Delta}
\frac{\dd x}{x}\Biggl[\biggl(1-x+\frac{x^2}{2}\biggr) \nonumber \\ &\times&
\left(L - 1 + \ln\frac{\theta_0^2}{4} + 2\ln(1-x)\right)
+ \frac{x^2}{2}\Biggr].   \label{dscd}
\end{eqnarray}
Note that there is an asymmetry between the contributions due to the emission
along the directions of the (initial or final) electron and
the ones due to production of collinear photons along the positron
momenta. The symmetry was broken when we decided to write a differential
cross--section with respect to the electron scattering angles
($\dd\Omega_- = \dd\cos(\widehat{\vecc{p}_-\vecc{p}}_-')\;\dd\phi)$.
After an integration over a symmetrical angular acceptance the
contributions would become equal.
Compensating terms are to be extracted from the
Eqs.(\ref{dsa},\ref{dsb},\ref{dscd}) by omitting the terms
proportional to $(L-1)$.

\section*{Appendix~3}

The following expression could be used instead of $WT$
in Eq.~(\ref{bhabha}) for more precise definition
of the contribution due to non--collinear hard photon emission in
large-angle Bhabha scattering:
\begin{eqnarray} \label{wtp}
(WT)_{\Pi} &=& \frac{(SS)}{|1-\Pi(s)|^2s\chi_-'\chi_+'}
+ \frac{(S_1S_1)}{|1-\Pi(s_1)|^2s_1\chi_-\chi_+}
- \frac{(TT)}{|1-\Pi(t)|^2t\chi_+\chi_+'} \\
&-& \frac{(T_1T_1)}{|1-\Pi(t_1)|^2t_1\chi_-\chi_-'}
+ \mbox{Re}\;\Biggl\{
\frac{(TT_1)}{(1-\Pi(t))(1-\Pi(t_1))^*
              tt_1\chi_-\chi_-'\chi_+\chi_+'} \nonumber \\
&-& \frac{(SS_1)}{(1-\Pi(s))(1-\Pi(s_1))^*
                            ss_1\chi_-\chi_-'\chi_+\chi_+'} \nonumber \\
&+& \frac{(TS)}{(1-\Pi(t))(1-\Pi(s))^*ts\chi_-'\chi_+\chi_+'}
+ \frac{(T_1S_1)}{(1-\Pi(t_1))(1-\Pi(s_1))^*
                         t_1s_1\chi_-\chi_-'\chi_+} \nonumber \\ \nonumber
&-& \frac{(T_1S)}{(1-\Pi(t_1))(1-\Pi(s))^*
                              t_1s\chi_-\chi_-'\chi_+'}
- \frac{(TS_1)}{(1-\Pi(t))(1-\Pi(s_1))^*ts_1\chi_-\chi_+\chi_+'}\Biggr\},
\end{eqnarray}
where
\begin{eqnarray*}
(SS)&=&(S_1S_1) = t^2 + t_1^2 + u^2 + u_1^2, \qquad
(TT)=(T_1T_1) = s^2 + s_1^2 + u^2 + u_1^2,  \\
(SS_1)&=& (t^2 + t_1^2 + u^2 + u_1^2)
( t\chi_+\chi_+' + t_1\chi_-\chi_-' - u\chi_+\chi_-' - u_1\chi_-\chi_+'), \\
(TT_1) &=& (s^2 + s_1^2 + u^2 + u_1^2)
( u\chi_+\chi_-' + u_1\chi_-\chi_+' + s\chi_-'\chi_+' + s_1\chi_-\chi_+), \\
(TS) &=& - \frac{1}{2}(u^2+u_1^2)\left[s(t+s_1)+t(s+t_1)-uu_1\right], \\
(TS_1) &=& - \frac{1}{2}(u^2+u_1^2)\left[t(s_1+t_1)+s_1(s+t)-uu_1\right], \\
(T_1S) &=& \frac{1}{2}(u^2+u_1^2)\left[t_1(s+t)+s(s_1+t_1)-uu_1\right], \\
(T_1S_1) &=& \frac{1}{2}(u^2+u_1^2)\left[s_1(s+t_1)+t_1(s_1+t)-uu_1\right].
\end{eqnarray*}
We checked analytically that for the {\em switched off}
vacuum polarization the above formula is equivalent
to the multiplication $WT$ in Eq.~(\ref{bere}):
\begin{eqnarray}
(WT)_{\Pi}|_{\Pi=0} = WT.
\end{eqnarray}

In the compensating terms, entering into Eq.~(\ref{bhabha}) the
vacuum polarization corrections, have to be inserted also. That can
be easily done starting with Eq.~(\ref{bhcoll}). We get
\begin{eqnarray}
\frac{\dd\sigma^{e^+e^-\to e^+e^-(\gamma)}_{\mathrm{comp}}}{\dd\Omega_-} &=&
\frac{\alpha}{\pi}
\int\limits_{\Delta}^{1}\frac{\dd x}{x}\; \Biggl\{
\biggl[\left(1-x+\frac{x^2}{2}\right)
\ln\frac{\theta_0^2(1-x)^2}{4}
+ \frac{x^2}{2}\biggr] \nonumber \\ &\times&  \nonumber
\frac{\dd\tilde{\sigma}_{0}(1,1)}{\dd\Omega_-}
\left(1+\frac{1}{(1-x)^2}\right)
+ \biggl[\left(1-x+\frac{x^2}{2}\right)
\ln\frac{\theta_0^2}{4} + \frac{x^2}{2}\biggr]
\\ &\times& \nonumber
\left[ \frac{\dd\tilde{\sigma}_{0}(1-x,1)}{\dd\Omega_-}
+ \frac{\dd\tilde{\sigma}_{0}(1,1-x)}{\dd\Omega_-} \right]
\Biggr\} \\
&-& \frac{\dd\tilde{\sigma}_{0}(1,1)}{\dd\Omega_-}
\frac{8\alpha}{\pi}\ln(\mbox{ctg}\frac{\theta}{2})
\ln\frac{\Delta\eps}{\eps}\;\dd\Omega_- \, .
\end{eqnarray}

We note that quantity $R$ (see Eq.(\ref{rmu}))
in the ultra--relativistic limit $s\gg m_{\mu}^2$
can be derived from Eq.(\ref{wtp}). One has to omit there
all terms except the ones proportional to $(SS)$, $(SS_1)$,
$(S_1S_1)$ and divide by 4.


\newpage

\begin{figure}[.3cm]
\end{figure}

\input epsf

\begin{figure}[t]
\begin{center}
\epsfbox[30 30 60 60]{e051a.eps}
\end{center}
\caption{The differential forward--backward asymmetry
for $e^+e^-\to \mu^+\mu^-$. The parameters are
$\varepsilon=0.51$~GeV, $\Psi_0=10^{\circ}$,
$\varepsilon_{\mathrm{th}}=0.2\varepsilon$.
}
\label{Fig1}
\end{figure}

\begin{figure}[1.3cm]
\end{figure}

\begin{figure}[t]
\begin{center}
\epsfbox[30 30 60 60]{e155a.eps}
\end{center}
\caption{The differential forward--backward asymmetry
for $e^+e^-\to \mu^+\mu^-$. Beam energy $\varepsilon=1.55$~GeV,
other parameters the same as in Fig.~1.
}
\label{Fig2}
\end{figure}

\end{document}